\documentclass[%
 onecolumn,  
superscriptaddress,
preprint,
nofootinbib,
 aps,
]{revtex4-2}

\usepackage{geometry}
\usepackage{blkarray, bigstrut}
\usepackage{lastpage} 
\usepackage[colorinlistoftodos]{todonotes}

\newgeometry{left=2cm,right=2cm,top=2cm,bottom=2cm}
\usepackage[english]{babel}

\usepackage{capt-of}
\usepackage{float}
\usepackage{graphicx}

\usepackage[acronym]{glossaries}

\usepackage{dcolumn}

\usepackage[breaklinks=true,colorlinks=true,linkcolor=blue,urlcolor=blue,citecolor=blue]{hyperref}

\usepackage[toc,page]{appendix}

\usepackage{enumitem}

\usepackage{amssymb}

\usepackage{mathtools}
\usepackage{braket}

\numberwithin{equation}{section}

\DeclarePairedDelimiter{\abs}{\lvert}{\rvert}
\DeclarePairedDelimiter{\norm}{\lVert}{\rVert}

\usepackage{amsthm}
\usepackage{amsmath}

\usepackage{bm}
\usepackage{bbold}

\theoremstyle{plain}

\theoremstyle{definition}

\newtheorem*{example*}{Example}

\usepackage{tcolorbox}

\allowdisplaybreaks

\usepackage{empheq}

\usepackage{tikz}

\usepackage{listings}

\usepackage{tikz}
\usetikzlibrary{quantikz}

\usepackage[normalem]{ulem}

\begin{document}

\newcommand{\rev}[1]{\color{red}{#1}\color{black}}

\title{A Near-Term Quantum Algorithm for Computing Molecular and Materials Properties based on Recursive Variational Series Methods }

\author{Phillip W. K. Jensen}
 \email{phillip.kastberg@gmail.com}
\affiliation{Chemical Physics Theory Group, Department of Chemistry, University of Toronto, Toronto, Ontario M5G 1Z8, Canada}
\affiliation{Zapata Computing Inc., 100 Federal Street, Boston, MA 02110, USA}

\author{Peter D. Johnson}
\email{peter@zapatacomputing.com}
\affiliation{Zapata Computing Inc., 100 Federal Street, Boston, MA 02110, USA}

\author{Alexander A. Kunitsa}
\email{aakunitsa@gmail.com}
\affiliation{Zapata Computing Inc., 100 Federal Street, Boston, MA 02110, USA}

\date{\today}

\begin{abstract}

Determining the properties of molecules and materials is one of the premier applications of quantum computing. A major question in the field is: how might we use imperfect near-term quantum computers to solve problems of practical value? We propose a quantum algorithm to estimate the properties of molecules using near-term quantum devices. The method is a recursive variational series estimation method, where we expand an operator of interest in terms of Chebyshev polynomials, and evaluate each term in the expansion using a variational quantum algorithm. We test our method by computing the one-particle Green's function in the energy domain and the autocorrelation function in the time domain.

\end{abstract}

\maketitle

\newacronym{rvse-c}{RVSE-C}{recursive variational series estimatiom}
\newacronym{gf}{GF}{Green's function}
\newcommand\NormChe{\overline{\chi}}
\newcommand\nmea{S}

\newcommand{\ak}[1]{\color{blue}{AK: #1}\color{black}}
\newcommand{\pj}[1]{\color{magenta}{PJ:#1}\color{black}}
\newcommand{\pk}[1]{\color{red}{PK:#1}\color{black}}






\section{Introduction}



For quantum computers of the distant future, powerful quantum algorithms have been developed for solving problems in quantum chemistry and materials.
These problems include estimating ground~\cite{lee_even_2021,su_fault-tolerant_2021} and excited state energies~\cite{bauman_toward_2021, jensen_quantum_2020}, molecular gradients ~\cite{obrien_efficient_2021}, and many others. 
However, implementing the quantum algorithms for problem instances of interest requires large-scale fault-tolerant quantum computers.
What are the prospects for solving such problems with quantum computers before such large-scale devices are available?
With the recent advances in quantum hardware, researchers have been investigating methods for using near-term quantum computers to solve problems of industry value.
For the area of quantum chemistry and materials, much of this work has focused on the task of ground state energy estimation, with the primary method being the variational quantum eigensolver (VQE) algorithm~\cite{mcclean_theory_2016, peruzzo_variational_2014}. 
Yet, for many problems of industrial value, properties beyond the ground state energy must be computed. {\color{black} To address this, the original VQE scheme was extended in two ways. First, it was combined with orthogonality constraints~\cite{higgott_variational_2019,ryabinkin_constrained_2019,yen_exact_2019} and quantum subspace expansion (QSE) techniques~\cite{mcclean_hybrid_2017,ollitrault_quantum_2020} to enable systematic exploration of the low-lying excited states. Second, finite-difference~\cite{obrien_calculating_2019} and analytical derivative methods~\cite{mitarai_theory_2020} were developed for the VQE energy functional to evaluate static molecular properties such as forces, vibrational frequencies, IR, and Raman intensities, magnetic exchange and shielding constants, to name a few. The resource requirements for such calculations have not been studied in detail. All of them, however, rely on the standard VQE observable estimation routines and are likely to suffer from the same measurement bottleneck~\cite{gonthier_identifying_2020}. Additionally, analytical derivative methods scale polynomially with the number of circuit parameters making them impractical for higher-order properties beyond relatively compact VQE ansatze. At a more fundamental level, all of the VQE-based algorithms share a common limitation: they require explicit preparation of target wave functions (either exact or approximate) that are then used to evaluate physical observables. Although straightforward to implement, such approach is inefficient compared to other methods that can directly provide relevant quantities, such as correlation functions. More versatile quantum algorithms have been introduced in the context of linear response~\cite{mcweeny_methods_1992} and Green's function~\cite{Fetter2003,Schirmer2018} formalism. Most of them rely on fault-tolerant computational subroutines~\cite{tong_fast_2021,cai_quantum_2020,keen_quantum_2021,kosugi_construction_2020} making their resource requirements prohibitive for implementation on noisy intermediate-scale quantum (NISQ) devices. This leaves open the question of whether efficient property estimation is feasible on the near-term quantum computers. Given that this is an important, though relatively underdeveloped area, it is valuable to explore novel approaches to solving these problems. Specifically, it is worthwhile to investigate alternative heuristic methods for estimating properties.
Following the same design principles as their fault-tolerant counterparts, most of the known NISQ algorithms for property estimation were derived by replacing demanding computational subroutines with their near-term versions. For example, Variational Quantum Simulation~\cite{yuan_theory_2019} and Quantum Imaginary Time Evolution~\cite{motta_determining_2020} were used to simulate quantum dynamics and compute correlation functions in real~\cite{endo_calculation_2020} and imaginary time~\cite{sun_quantum_2021,tazhigulov_simulating_2022,sakurai_hybrid_2021}, respectively. Similarly, variational linear system solvers were applied to response function calculations~\cite{chen_variational_2021}. QSE and quantum Equation-of-Motion~\cite{ollitrault_quantum_2020} methods were adapted for Green's function evaluation~\cite{rizzo_one-particle_2022,endo_calculation_2020}.
Most of these methods are fairly specific to the properties being estimated. Moreover, their performance is sensitive to the underlying assumptions and has not been thoroughly analyzed beyond small-scale demonstrations on photonic and superconducting quantum devices~\cite{zhu_calculating_2021,huang_quantum_2022,keen_quantum-classical_2020}.
With this motivation, our work explores the following question: How can we use near-term quantum computers to estimate molecular spectra and more general properties of the approximate ground and excited states? 

General dynamical properties can be expressed as spectral functions in the energy (or frequency) domain.  It is often sufficient to sketch a spectral function of interest to make predictions about certain properties. Viewed from this perspective, the problem can be tackled with classical approximation techniques such as kernel polynomial method~\cite{weisse_kernel_2006} (KPM). KPM reconstructs spectral functions based on their moments in a suitable basis. For improved numerical stability, they are commonly used in conjunction with Chebyshev expansion~\cite{ferreira_critical_2015, sobczyk_spectral_2021}. This enables moment calculations via an iterative process that effectively applies a function of the Hamiltonian $\hat{H}$, $\hat{T}_k(\hat{H})$, to an arbitrary initial state, where $T_k$ is the k-th order Chebyshev polynomial of the first kind. This process is amenable to both classical and quantum computation. The main obstacle in the latter case is implementing the non-unitary operator $\hat{T}_k(\hat{H})$, which can be accomplished, for example, with block-encodings~\cite{rall_quantum_2020} and qubitization~\cite{low_hamiltonian_2019}. Alternatively, one can circumvent the problem by switching to a Fourier basis, as shown in Ref.~\cite{wang_kernel_2022}. Lanczos recursion is closely related to KPM, but instead of using moment expansion, it performs iterative construction of the continued fraction, representing a spectral function. Recently, it has been studied in the context of fault-tolerant~\cite{baker_lanczos_2021} and near-term quantum computing as a tool for Green's function calculations~\cite{jamet_krylov_2021}. 
}

In this work, we introduce a flexible method for using low depth quantum circuits to estimate expectation values of general functions of a Hamiltonian with respect to an input state. It is inspired by KPM and has the same objective as the property estimation algorithm of Rall~\cite{rall_quantum_2020} while using far less circuit depth. In our method, a function of interest is approximated by a series expansion, and we recursively train a parameterized circuit to approximate states proportional to the series terms applied to the input state.
Accordingly, we refer to this method as \emph{recursive variational series estimation} (RVSE). The method uses Hadamard tests to compute the overlaps between states output from short-depth quantum circuits, making this approach suitable for near-term quantum computation. We assess the performance of the method for two different applications with simulations that include statistical (sampling) noise, investigating the trade-off between accuracy and error. Additionally, we model the buildup of error due to the recursive nature of the algorithm. Example applications of RSVE include estimating the spectrum for Green's function methods and calculating autocorrelation functions. 


The paper is structured as follows: In Section~\ref{sec:chebyshev_method}, we review the Chebyshev expansion method and some of its applications in quantum physics. Section~\ref{sec:rvse} describes the details of RVSE. Computational tests of the algorithm are presented in Section~\ref{sec:compu_tests}, showcasing its performance for smooth and singular functions in the presence of sampling noise. We conclude with a discussion of open questions and future research directions in Section~\ref{sec:outlook_and_conclusions}.

\section{The Chebyshev Method}
\label{sec:chebyshev_method}


The Chebyshev polynomials of the first kind are a class of orthogonal polynomials related to the cosine function, $T_k(\omega) \equiv \cos(k \arccos\omega)$, such that any smooth function in the interval $\omega \in [-1,1]$ can be expanded in a Chebyshev series: 

\begin{align}
f(\omega)  = \sum^{\infty}_{k =0}   c_k T_k(\omega)
\label{eq:chebyshev_series}
\end{align}
where the expansion coefficients are given by

\begin{align}
c_k(\omega) = \frac{2-\delta_{k0}}{\pi} \int^{1}_{-1} \frac{ f(\omega) T_k(\omega)}{\sqrt{1-\omega^2}}d\omega.  \label{eq:expand_coefs}
\end{align}
The Chebyshev polynomials satisfy the following recurrence relation 
\begin{align}
T_0(\omega) &= 1 \nonumber \\
T_1(\omega) &= \omega \nonumber \\
T_k(\omega) &= 2\omega T_{k-1}(\omega) - T_{k-2}(\omega). \label{eq:recursion_relation}
\end{align}
Note that only the last two polynomials are required at each recursion step. For continuous functions $f(\omega)$, the expansion~\eqref{eq:chebyshev_series} converges uniformly on $[-1,1]$~\cite{rivlin_chebyshev_2020}. In practice, it can also be applied to singular functions in conjunction with regularization techniques described in~\cite{weisse_kernel_2006}. The truncation of the infinite series in Eq. \eqref{eq:chebyshev_series} leads to fluctuations --- also known as Gibbs oscillations --- near the points where the function is discontinuous~\cite{weisse_kernel_2006}. It is possible to dampen these fluctuations by modifying the Chebyshev series using suitably chosen kernels. To see this, we rewrite the expansion \eqref{eq:chebyshev_series} as follows~\cite{weisse_kernel_2006}


\begin{align}
f(\omega) = \frac{1}{\pi \sqrt{1-\omega^2}} \bigg[ \mu_0    + 2 \sum^\infty_{k = 1} \mu_k T_k(\omega) \bigg] \label{eq:chebyshev_series_kpm}
\end{align}
with coefficients or moments 

\begin{align}
\mu_k = \int^{1}_{-1} f(\omega) T_k(\omega) d\omega. 
\end{align}
The kernel polynomial method maps the function \eqref{eq:chebyshev_series_kpm} to $ f_{\text{KMP}}(\omega)$ by modifying the moments $\mu_k \rightarrow g_k \mu_k$. In this work, we consider the simplest kernel: the Dirichlet kernel by setting $g_k = 1$, which yields the truncated series in Eq. \eqref{eq:chebyshev_series} and leads to the disadvantages mentioned earlier, but our method can be extended to other kernels. The concept of Chebyshev expansions can be further extended to operators, i.e., any operator function  $\hat{F}(\hat{H}|E)$ can be expressed in terms of the Chebyshev operators $\hat{T}_k(\hat{H})$ as long all of the eigenvalues of $\hat{H}$ lie between [-1,1]. If the condition does not hold, the original operator needs to be re-scaled via a linear transformation:

\begin{align}
\hat{H}_{\text{sc}} = (\hat{H} - H^+)/H^-, \label{eq:scaled_hamil}
\end{align}
where $H^{\pm} = (E_{\text{max}}  \pm E_{\text{min}} ) / 2$, and $E_{\text{min}} $ and $E_{\text{max}}$ denote minimum and maximum eigenvalues of $\hat{H}$, respectively. Computing the exact values of $E_{\text{max}}$ and $E_{\text{min}} $ is expensive, and we can loosen this restriction by choosing instead $\tilde{E}_{\text{max}}$ and $\tilde{E}_{\text{min}}$ such that $E_{\text{max}} < \tilde{E}_{\text{max}}$ and $E_{\text{min}} > \tilde{E}_{\text{min}}$ which ensure all the eigenvalues lie in $(-1,1)$. Another option is to scale the Hamiltonian as 

\begin{align}
\hat{H}_{\text{sc}} = \frac{1}{\lambda} \hat{H} && \norm{\hat{H}}_2 \leq \lambda = \sum^L_{j} \abs{h_j}, \label{eq:scaled_hamil_1}
\end{align}
where $\norm{\hat{H}}_2$ is the spectral norm and  $\lambda$ is the $L_1$-norm.  We use the method in \eqref{eq:scaled_hamil} in  Section \ref{sec:compu_tests}, where  the minimum and maximum eigenvalues of $\hat{H}$ can easily be obtained for the molecular system considered here.


Chebyshev expansion techniques found multiple applications in computational condensed matter physics~\cite{weisse_kernel_2006} and quantum chemistry~\cite{chen_evolution_1996,chen_chebyshev_1999, tal-ezer_accurate_1984, zhu_orthogonal_1994, chen_discrete_1998}. For example, it was shown by Tal-Ezer and Kolsloff~\cite{tal-ezer_accurate_1984} that expanding the time-evolution operator $\exp(-i\hat{H}t)$ in a \emph{K}-term Chebyshev series converges exponentially with $K$ for $K> t$ where \emph{t} is the evolution time. Thus, long expansions are not necessary, which is computationally efficient. Other examples are the imaginary time-evolution operator  $\exp(-\tau \hat{H})$ for which Chebyshev expansion converges even faster~\cite{kosloff_direct_1986}, and the resolvent  $(E-\hat{H})^{-1}$, which is useful to compute Green's functions and related quantities in the energy domain~\cite{chen_evolution_1996, chen_chebyshev_1999, zhu_orthogonal_1994}. 

One of the advantages of expanding the operator of interest in a Chebyshev series is that storing the Hamiltonian matrix is not necessary. For example, obtaining the inverse of the Hamiltonian matrix requires matrix manipulations hence storing, but the problem can be reduced to matrix-vector multiplication of the form $H\vec{c}$ using the Chebyshev expansion. For a sparse matrix, the number of non-zero elements scale as $\mathcal{O}(D)$, where \emph{D} is the dimension of the matrix, and  the calculation of a \emph{K}-term Chebyshev expansion, therefore, requires $\mathcal{O}(KD)$ operations and time. The matrix-vector multiplication is, however, not a unique property since in quantum chemistry one rarely stores the Hamiltonian matrix. For example, diagonalization algorithms such as the Davidson method~\cite{davidson_iterative_1975} require matrix-vector multiplication only. A more unique property of the Chebyshev method is its fast convergence for certain problem instances which we will investigate in Section \ref{sec:compu_tests}. Why the need for a quantum computer if matrix-vector multiplication can already be done efficiently on a classical device? Consider the molecular electronic Hamiltonian in the canonical Hartree-Fock spin-orbital basis, then its Hamiltonian matrix representation is indeed sparse (see for instance~\cite{helgaker_molecular_2000}), and it therefore only requires $\mathcal{O}(KD)$  operations. However, the number of non-zero elements still has a steep scaling  $D=\mathcal{O}(M^{N})$, where \emph{M} is the number of spin orbitals and \emph{N} is the number of electrons, making even matrix-vector multiplication intractable~\cite{helgaker_molecular_2000}.

In this work, we explore recursive methods, specifically the Chebyshev method, on a digital quantum computer. The quantum algorithm proposed here overcomes the bottleneck of matrix-vector multiplication, but, as we will see, other problems arise such as sampling noise and circuit optimization. In the following, we will assume all quantum states required to implement RVSE can be prepared efficiently, leaving a more detailed exploration of this issue for future research.




\section{Recursive Variational Series Estimation Method}
\label{sec:rvse}

\begin{figure}[t] 
\centering  
\includegraphics[width=0.9\textwidth]{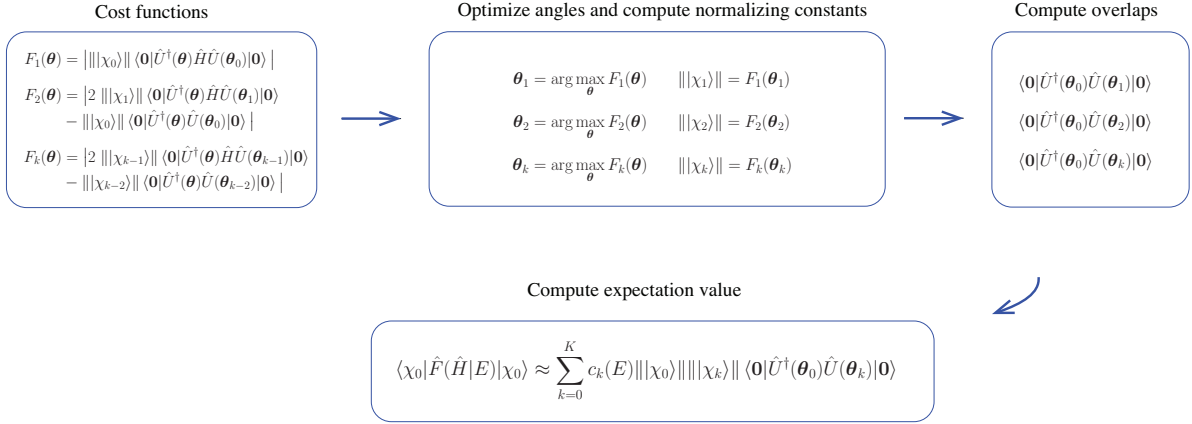}
\caption{Schematic representation of the recursive variational series estimation (RVSE) method. The coefficients $c_k(E)$ are computed on a classical computer, and the wave function overlaps on a quantum computer.  }
\label{fig:alg}
\end{figure}

In this section, we introduce the recursive variational series estimation algorithm. The first step of the algorithm is to expand the target operator in the Chebyshev basis. Starting with an initial state $\ket{\chi_{0}}$, the action of the operator $\hat{F}(\hat{H}|E)$ on the initial state can be evaluated by

\begin{align}
 \hat{F}(\hat{H}|E)\ket{\chi_0} \approx \sum^K_{k = 0}  c_k(E) \ket{\chi_k} = \sum^K_{k = 0}   c_k(E) \bigg( 2 \hat{H} \ket{\chi_{k-1}} - \ket{\chi_{k-2}} \bigg), \label{eq_f}
\end{align}
where $\ket{\chi_k} \equiv  \hat{T}_k(\hat{H}) \ket{\chi_0}$, and the expansion coefficients can be obtained from Eq. \eqref{eq:expand_coefs}. The Hamiltonian is scaled such that all the eigenvalues lie in $(-1,1)$ (we shall omit the label `sc' to simplify the notation), e.g., using the protocols in \eqref{eq:scaled_hamil} or \eqref{eq:scaled_hamil_1}.  We can estimate the expectation value of the function $\hat{F}(\hat{H}|E)$ as

\begin{align}
\braket{\chi_0| \hat{F}(\hat{H}|E)|\chi_0} \approx \sum^K_{k = 0}  c_k(E) \braket{\chi_0|\chi_k} = \sum^K_{k = 0}   c_k(E) \bigg( 2 \braket{\chi_0|\hat{H}|\chi_{k-1}} - \braket{\chi_0|\chi_{k-2}} \bigg), \label{eq:expectation_value}
\end{align}
which requires the overlaps of the form $\braket{\chi_0| \hat{H}|\chi_{k-1}}$ and $\braket{\chi_0|\chi_{k-2}}$. In general, the Chebyshev operator $ \hat{T}_k(\hat{H})$ is not unitary, and the resulting Chebyshev state $\ket{\chi_k}$ is therefore unnormalized. For that reason, we cannot directly prepare the Chebyshev states on a quantum computer. We can, however, represent a normalized state $\ket{\NormChe_k} \equiv \ket{\chi_k}/\norm{\ket{\chi_k}}$ using a parametrized quantum circuit so that $\ket{\NormChe_k} =\hat{U}(\boldsymbol{\theta}_k) \ket{0}$ where $\hat{U}(\boldsymbol{\theta}_k)$ is a unitary operator and $\boldsymbol{\theta}_k$ is a vector of real parameters. To prepare $\ket{\NormChe_k}$, we define an overlap between the true and normalized states as

\begin{align}
F_k(\boldsymbol{\theta}) &=  \left\lvert \bra{\boldsymbol{0}}\hat{U}^\dagger(\boldsymbol{\theta}) \ket{\chi_k}  \right\lvert  \\
&=   \norm{\ket{\chi_{k}}} \left\lvert \bra{\boldsymbol{0}}\hat{U}^\dagger(\boldsymbol{\theta}) \ket{\NormChe_k}  \right\lvert  \\
&=   \left\lvert2 \norm{\ket{\chi_{k-1}}} \bra{\boldsymbol{0}} \hat{U}^\dagger(\boldsymbol{\theta}) \hat{H}\ket{\NormChe_{k-1}} - \norm{\ket{\chi_{k-2}}}  \bra{\boldsymbol{0}} \hat{U}^\dagger(\boldsymbol{\theta})\ket{\NormChe_{k-2}}  \right\lvert  \\
&= \left\lvert2 \norm{\ket{\chi_{k-1}}} \sum^L_{j=1} h_j\braket{\boldsymbol{0}|\hat{U}^\dagger(\boldsymbol{\theta})\hat{P}_j \hat{U}(\boldsymbol{\theta}_{k-1})|\boldsymbol{0}} -   \norm{\ket{\chi_{k-2}}} \braket{\boldsymbol{0}|\hat{U}^\dagger(\boldsymbol{\theta}) \hat{U}(\boldsymbol{\theta}_{k-2})|\boldsymbol{0}}  \right\lvert, \label{eq:overlap:func}
\end{align}
where $\ket{\boldsymbol{0}} \equiv \ket{0 \hdots 0}$, and

\begin{align}
\hat{H} =  \sum^L_{j = 1} h_j \hat{P}_j \label{eq_qu_hamil}
\end{align}
is the Hamiltonian decomposed as a linear combination of unitaries $\hat{P}_j$ with scaled Hamiltonian coefficients $h_j \in \mathbb{R}$. A similar strategy was proposed in Ref.~\cite{jamet_krylov_2021} for preparing the Lanczos states on a quantum computer.  We note that $F_k(\boldsymbol{\theta}) \leq  \norm{\ket{\chi_{k}}} $, and the maximum is achieved for the set of parameters $\boldsymbol{\theta}_k$ making $\ket{\NormChe_k}$ aligned with $\ket{\chi_k}$ such that $F_k(\boldsymbol{\theta}_k) =  \norm{\ket{\chi_{k}}}$. The optimization is carried out by maximizing the cost function in Eq. \eqref{eq:overlap:func}. In an ideal situation the optimized cost function results in the optimal angles $\boldsymbol{\theta}_k$, and the value of the cost function equals the normalizing constant $ \norm{\ket{\chi_{k}}}$. We can then estimate the expectation value in terms of the normalizing constants as

\begin{align}
\braket{\chi_0|\hat{F}(\hat{H}|E)|\chi_0} 
&\approx   \sum^K_{k = 0} c_k(E) \norm{\ket{\chi_0}} \norm{\ket{\chi_k}} \braket{\textbf{0}|\hat{U}^\dagger(\boldsymbol{\theta}_0)\hat{U}(\boldsymbol{\theta}_k)|\textbf{0}}, \label{eq:final_overlap}
\end{align}
where the wave function overlaps $\braket{\textbf{0}|\hat{U}^\dagger(\boldsymbol{\theta}_0)\hat{U}(\boldsymbol{\theta}_k)|\textbf{0}}$ can be computed on a quantum computer, e.g., using the Hadamard test circuit~\cite{aharonov_polynomial_2009,mitarai_methodology_2010}. A schematic representation of the algorithm is shown in Fig. \ref{fig:alg}.

In practice, the optimization will be prone to errors due to, e.g., sampling noise, decoherence, and the presence of local maxima, resulting in approximate angles and normalizing constants. Furthermore, recursive algorithms depend on previous steps in the algorithm (Eq. \eqref{eq:recursion_relation}) thus errors can propagate through the recursion and disrupt the computation. This is less of a problem for a classical computer because the matrix elements are evaluated with high precision. In this work, we assume the ansatz is expressive enough to reach the target states at each iteration, and the optimization algorithm is able to locate the global maximum, i.e., we obtain the optimal angles $\boldsymbol{\theta}_k$ at each iteration. Note that for the ansatz to be expressive enough to reach the target states at each iteration, we anticipate that the number of parameters in such an ansatz increases with \emph{k}. We test our algorithm in the presence of sampling noise when computing the cost function in Eq. \eqref{eq:overlap:func}, given the optimal angles $F_k(\boldsymbol{\theta}_k)$, by adding errors into the wave function overlaps. Recall that the $k^{\text{th}}$ Chebyshev polynomial depends on the previous two, hence even small errors can accumulate through the recursion and disrupt the computation. In Appendix \ref{app_cheby}, we show that the normalizing constants in the presence of sampling noise, and in the large sample limit, can be separated into their exact and noisy components as 

\begin{align}
 \norm{\ket{\chi_k}}_{\text{noisy}} =   \norm{\ket{\chi_k}} +\epsilon_k \label{eq_k_norm}
\end{align}
where $ \norm{\ket{\chi_k}}_{\text{noisy}}$ denotes the expectation value of the cost function estimator,  Eq.~\eqref{eq_app:noisy_k}, and the error $\epsilon_k$ depends on the previous errors $\epsilon_{k-1}$ and $\epsilon_{k-2}$, thus error can potentially propagate.  We therefore want to test how sensitive the RVSE algorithm is to sampling noise. Other sources of error, e.g., device noise, are omitted in this work since we consider robustness with respect to sampling noise an important prerequisite for the algorithm performance.

In the next section, we test the RVSE  performance in the presence of sampling noise when computing the one-particle Green's function and auto-correlation function for molecular hydrogen in the 6-31G basis~\cite{hehre_selfconsistent_1972}. Green's function is a central quantity in various fields, e.g., quantum transport theory~\cite{meir_landauer_1992}, propagator methods~\cite{oddershede_polarization_1984}, photoemission spectroscopy, making it an important target for investigation. The time-dependent auto-correlation function requires Hamiltonian simulation, and we use the RVSE method to implement the time-evolution operator. For Hamiltonian simulation, our method is similar to the qubitization protocol in Ref. \cite{low_hamiltonian_2019} 
, but our approach is based on a quantum variational algorithm which is not the case with qubitization. 

\section{Computational Tests}
\label{sec:compu_tests}

\subsection{The One-Particle Green's Function}
\label{sec:gf}

In this example, we consider the retarded one-particle Green's function (\acrshort{gf}) at zero temperature  defined as $G_{ij}^r(t)=-i\theta(t)\braket{\{\hat{a}_i(t),\hat{a}^\dagger_j(0)\}}$, where $\theta(t)$ is the Heaviside step function, $\hat{a}_i(t) = e^{i \hat{H} t} \hat{a}_i e^{-i \hat{H} t}$ is the Heisenberg representation of the fermion annihilation operator $\hat{a}_i$, and $\braket{\hdots} = \bra{E_0}\hdots \ket{E_0}$ denotes the expectation value with respect to the \emph{N}-particle ground state of $\hat{H}$.  It is assumed that we have already solved for the ground state, which itself is a highly non-trivial problem (see for instance~\cite{ogorman_intractability_2021,anand_quantum_2022, cao_quantum_2019}). In the past years, several quantum algorithms have been proposed to compute \acrshort{gf}s on a quantum computer~\cite{baker_lanczos_2021, tong_fast_2021, kosugi_construction_2020,endo_calculation_2020, jamet_krylov_2021, rizzo_one-particle_2022}. The method most similar to ours is the Lanczos recursion method~\cite{jamet_krylov_2021}, which variationally prepares the Lanczos states on a quantum computer and computes the one-particle \acrshort{gf} based on the continued fraction representation. 

It is convenient to express the \acrshort{gf} in the energy domain obtained by a Fourier transform of $G_{ij}^r(t)$: 

\begin{align} 
G^r_{ij}(E) = \braket{\Phi_i|\frac{1}{z^+ -\hat{H} }  |\Phi^j } -  \braket{\Phi^j|\frac{1}{z^- - \hat{H}}|\Phi_i }, \label{eq:compu_tests_retarded_gf}
\end{align}
where $z^{\pm} = \pm(E + i\eta) + E^{(N)}_{0}$ with $E^{(N)}_{0}$ as the ground state energy for the neutral system with \emph{N} particles, and $\ket{\Phi_i} \equiv a_i \ket{E_{0}}$ $(\ket{\Phi^j} \equiv a_j^\dagger \ket{E_{0}})$ is the electron removal (attachment) state. We will focus on the spectral function, defined as:

\begin{align}
A_{ij}(E) &= -\frac{1}{\pi} \text{Im}G^r_{ij}(E)
\label{eq:compu_tests_spectral_function}
\end{align}
The spectral function describes electron attachment and removal and has poles at electron affinities (EA) and ionization potentials (IP).  The width of the peaks is given by $\frac{1}{2} \eta$. The EA and IP, with respect to the ground state energy, are given by  $\text{EA}_a = E^{(N)}_0 - E^{(N+1)}_a$ and $\text{IP}_i =   E^{(N-1)}_i  - E^{(N)}_0$, respectively, and can be extracted from the spectral function. Note that the spectral function \eqref{eq:compu_tests_spectral_function} has poles at the negative values of EA and IP. 
In the limit   $\eta \rightarrow 0^+$, the spectral function converges to a weighted sum of delta functions, $\lim_{\eta \rightarrow 0^+} A_{ij}(E)  \rightarrow  \sum_k a_k^{(ij)} \delta( E + E^{(N)}_{0} - E_k^{(N+1)}) +\sum_l b_l^{(ij)} \delta( E - E^{(N)}_{0} + E_l^{(N-1)})$.

We expand the expectation value of the Green's operator $(z - \hat{H})^{-1}$ with respect to an arbitrary state $\ket{\chi_0}$ in the Chebyshev basis as follows~\cite{huang_general_1993,lipkowitz_recursive_2007}

\begin{align}
\bra{\chi_0}\frac{1}{z - \hat{H} }\ket{\chi_0}  &=  \frac{-i}{\sqrt{1-z^2}} \sum^{\infty}_{k = 0} (2 - \delta_{k0}) e^{-ik \arccos z}\braket{\chi_0|\hat{T}_k(\hat{H})|\chi_0}.   \label{eq:GO_chebyshev}
\end{align}
The overlaps $\braket{\chi_0|\hat{T}_k(\hat{H})|\chi_0} $ are then estimated on a quantum computer, as described in Section \ref{sec:rvse}. Note that the expansion \eqref{eq:GO_chebyshev} does not decay as \emph{k} increases and truncation at finite \emph{k} results in fluctuations, also known as Gibbs oscillations, near the poles. Consequently, even the last term contributes fully to the expansion making the spectral function a difficult target for the Chebyshev method. However, the fluctuation near the energy eigenvalues can be reduced using a different kernel which smoothes the function~\cite{weisse_kernel_2006}.

\begin{figure}[t] 
\centering  
\includegraphics[width=0.9\textwidth]{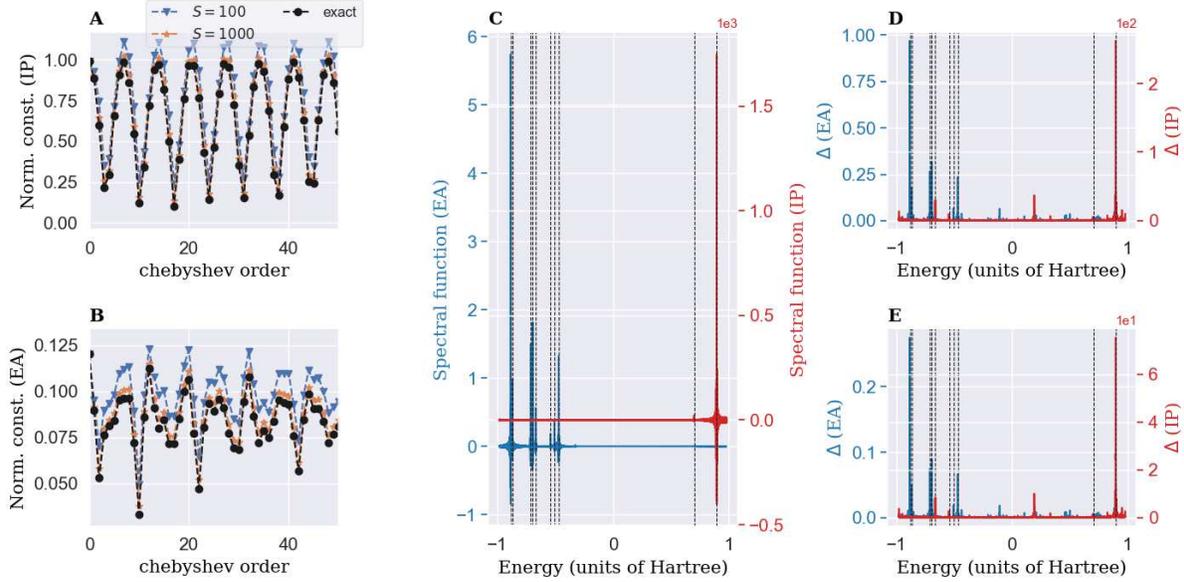}
\caption{ Spectral function for molecule Hydrogen in the 6-31G basis at the equilibrium bond length 0.74 \text{\r{A}} using the RVSE method. EA and IP on the y-axis indicate the first and second terms in Eq.~\eqref{eq_A_11}, respectively. The vertical dashed lines highlight the exact EA and IP energies. \emph{S} denotes the number of measurements used to evaluate the cost function. We set $S = 100$ and $S = 1000$ in  Figs. D and E, respectively.   }
\label{fig:spectral_func}
\end{figure}

We test the method by computing the diagonal element of the spectral function for molecular Hydrogen

\begin{align}
A_{11}(E)   = - \frac{1}{\pi} \text{Im}
\left(\braket{E_0| 
\hat{a}_1\frac{1}{E + i \eta + E^{(N)}_0 -\hat{H}} \hat{a}^\dagger_1 |E_0  }- \braket{E_0| 
\hat{a}^\dagger_1\frac{1}{-(E + i \eta) + E^{(N)}_0 -\hat{H}} \hat{a}_1 |E_0  } \right)\label{eq_A_11}
\end{align}
where the molecular orbital index starts from 0,  and $\hat{H}$ is the electronic Hamiltonian for molecular Hydrogen in the 6-31G basis at the equilibrium bond length 0.74 \text{\r{A}}. The Hamiltonian is scaled using the protocol in \eqref{eq:scaled_hamil} where $E_{\text{max}}$ and $E_{\text{min}}$ can easily be computed for this system. We set the initial state  to $\ket{\chi_0} = \hat{a}^\dagger \ket{E_0}$ in the Chebyshev expansion \eqref{eq:GO_chebyshev} to evaluate the first term in \eqref{eq_A_11} and $\ket{\chi_0} =\hat{a}_1 \ket{E_0}$ for the second term. The exact normalizing constants are shown in Figs. \ref{fig:spectral_func}~A and B for the first 50 terms in the Chebyshev expansion. Next, we tested our RVSE method in the presence of sampling noise using the noise model explained in Appendix \ref{app_cheby}. The noisy normalizing constants can be computed from Eq. \eqref{eq_k_norm}, and we want to observe the effect of $\epsilon_k$. Based on these simulations, it seems $\epsilon_k$ is not building up error, that is, error propagation does not have a significant impact for this example. In fact, the noisy normalizing constants seem to be only shifted from their true value, which can be corrected. Figure \eqref{fig:spectral_func}~C  shows the exact (noiseless) spectral function \eqref{eq_A_11}, setting $K=2000$  to obtain a smooth function and distinguish the tiny peaks corresponding to eigenvalues from background oscillation in the spectrum. The spectral function shows peaks at the EA and IP energies, as expected. In the presence of sampling noise, and in the large sample limit, the absolute error between the exact \emph{K}-term Chebyshev expansion and its noisy value is given by (Appendix \ref{subsec:exp})

\begin{align}
\Delta = \left| \sum^K_{k = 0} c_k(E) \norm{\ket{\chi_0}} \epsilon_k\braket{\textbf{0}|\hat{U}^\dagger(\boldsymbol{\theta}_0)\hat{U}(\boldsymbol{\theta}_k)|\textbf{0}}  \right|  \label{eq_abs_error} 
\end{align}
where $c_k(E) = \frac{-i}{\sqrt{1-E^2}} (2-\delta_{k0}) \exp(-ik\arccos(E))$ and $\epsilon_k$ is the error in the \emph{k}'th normalizing constant, Eq. \eqref{eq_k_norm}. Note that the error function \eqref{eq_abs_error} is almost identical to the Chebyshev expansion, $\norm{\ket{\chi_k}} \rightarrow \epsilon_k$, and since $\epsilon_k$ follows the same shape as $\norm{\ket{\chi_k}}$ (Fig. \ref{fig:spectral_func}~A and B), the error function is similar to the Chebyshev expansion, as observed in  Figs. \eqref{fig:spectral_func}~D and E. The error can be reduced by increasing the number of measurements \emph{S} to evaluate the cost functions, which decreases $\epsilon_k$.

\subsection{The Auto-Correlation Function}

The Chebyshev expansion of the time-evolution operator was worked out by Tal-Ezer and Kosloff~\cite{tal-ezer_accurate_1984}:

\begin{align}
e^{-i\hat{H}_{\text{sc}} t} &\approx  \sum^K_{k = 0} (2-\delta_{k0}) (-i)^k J_k(t) \hat{T}_k(\hat{H}_{\text{sc}}), \label{eq:time_evovl_che_1} 
\end{align}
where $J_k$ is the \emph{k}'th Bessel function of the first kind. The truncation error can be bounded as~\cite{berry_hamiltonian_2015}

\begin{align}
\epsilon = \norm{ \hat{U}_K-\hat{U}_{\infty}} \leq  \sum^\infty_{k = K+1}2 \abs{J_k(t)} \leq \frac{4 t^{K+1}}{2^{K+1} (K+1)!}  = \mathcal{O}( ( et/K)^K))\label{eq:time_evovl_error_bound}
\end{align}
where $\hat{U}_{\infty}$ is the exact unitary operation. Note that the error dramatically decreases for $K>t$ and in the limit of large \emph{K} converges to zero, and therefore only a small number of extra terms are needed beyond $K = t$. In fact, to ensure that the error is at most $\epsilon$, it suffices to take $K = \mathcal{O}(\abs{t} + \ln(1/\epsilon))$~\cite{berry_hamiltonian_2015}. The auto-correlation function can then be approximated as

\begin{align}
C(t) &= \braket{\Psi|e^{-i\hat{H}_{\text{sc}} t}|\Psi} \approx \sum^K_{k = 0} (2-\delta_{k0}) (-i)^k J_k(t) \braket{\Psi| \hat{T}_k(\hat{H}_{\text{sc}}) |\Psi}. \label{eq:auto_corr_cheby}
\end{align}
Compared to the spectral function in the previous section, the auto-correlation function converges much faster since $e^{- i E t}$ is a smooth function~\cite{boyd_rootfinding_2007}. It is known that the Chebyshev expansion leads to relatively poor convergence at the discontinuities or singularities due to increasing fluctuations, and the  spectral function is a weighted sum of sharp peaks. Hence it requires many terms in the expansion.



\begin{figure}[t] 
\centering  
\includegraphics[width=0.8\textwidth]{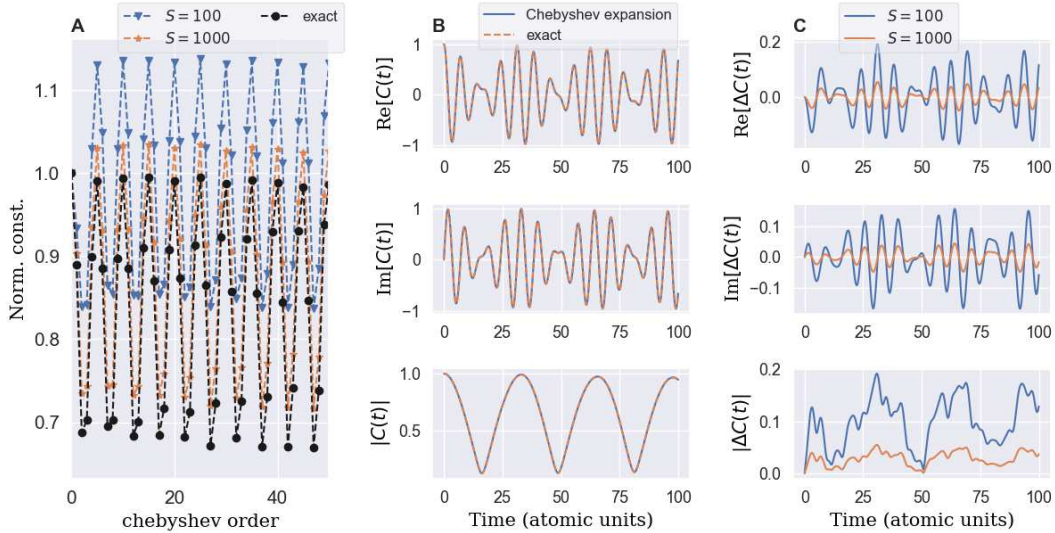}
\caption{ Auto-correlation function $C(t)$ in the time domain for molecular hydrogen using the electronic Hamiltonian in the 6-31G basis at equilibrium bond length 0.74 \text{\r{A}}.  }
\label{fig:auto_func}
\end{figure}

We test our method by computing the auto-correlation function in \eqref{eq:auto_corr_cheby} using the electronic molecular Hamiltonian for H$_\text{2}$ in the 6-31G basis at equilibrium bond length 0.74 \text{\r{A}}. We set the initial state to $\ket{\chi_0} =\ket{\Psi} = \frac{1}{\sqrt{2}} ( \ket{\phi_0 \phi_1} + \ket{\phi_2 \phi_3}) $, where $\ket{\phi_i \phi_j}$ indicates the spin orbitals \emph{i} and \emph{j} are occupied, which is a singlet state with non-zero overlap with the ground state. Figure \ref{fig:auto_func}~A shows the exact and noisy normalizing constants for the first 50 terms in the Chebyshev expansion. As for the spectral function, the effect of error propagating in $\epsilon_k$, Eq. \eqref{eq_k_norm}, is not significant but only shift the curve. Figure \ref{fig:auto_func}~B shows the exact auto-correlation function, $C(t) = \sum_{n} \abs{\braket{E_n|\Psi}}^2 e^{-iE_nt}$, which  consists of a mix of eigenstates each contributing a different period of $2\pi/\abs{E_n}$. We performed an exact diagonalization of the Hamiltonian to compute the reference auto-correlation function. We compare the exact auto-correlation function with its Chebyshev expansion \eqref{eq:auto_corr_cheby} using 120 terms in the expansion, and we observe the truncation error is negligible, when setting the maximal evolution time $t = 100$ a.u, as expected from Eq. \eqref{eq:time_evovl_error_bound}. We then tested our method in the presence of sampling noise. The absolute error is given in Eq. \eqref{eq_abs_error}, where $c_k(t) = \left( 2- \delta_{k0} \right)(-i)^k J_k(t)$ and $\epsilon_k$ is the error in the \emph{k}-th normalizing constant. We observe that the error function follows the same function as the auto-correlation function, where the error can be reduced by increasing the number of measurements to evaluate the cost function.

\section{Conclusion and Outlook}
\label{sec:outlook_and_conclusions}

RVSE is a mathematical framework for estimating expectation values of general operators. Its key component is a series expansion of a target operator in the Chebyshev polynomial basis which can be implemented variationally by introducing an ansatz for each term. RVSE is therefore inspired by the Kernel Polynomial Method used to compute spectral functions on classical computers~\cite{weisse_kernel_2006}, but is applicable in a more general context. Here, we focused on the problem of evaluating expectation values of operators represented as functions of the Hamiltonian. Specifically, we tested RVSE for Hamiltonian simulation and one-particle Green's function calculations, both of which are promising applications of near-term quantum computers. Without performing a detailed comparison with the state-of-the-art algorithms, 
we focused on assessing the noise resilience of RVSE and its resource efficiency through numerical simulations. Our implementation benefits from recursive properties of Chebyshev polynomials to avoid estimating increasingly high powers of the Hamiltonian operator $\braket{\hat{H}^{k}}$. This keeps the number of measurements approximately constant at each step of recursion, but introduces interdependence between the consecutive terms in the series allowing sampling noise amplification. 
Yet, 
our algorithm 
faithfully reproduces the overall shape of spectral function for the model example of H$_2$ molecule. Using the auto-correlation function as another example, we studied the interplay of the series truncation and sampling noise accrued while evaluating it. The Chebyshev basis is known to be efficient for representing the time evolution operator and has been previously used to derive optimized Hamiltonian simulation algorithms~\cite{low_optimal_2017}. RVSE adopts a similar approach offering an alternative to existing near-term methods such as Variational Quantum Simulation~\cite{higgott_variational_2019}
and related techniques~\cite{yuan_theory_2019,barison_efficient_2021}. Depending on simulation time, only a few terms of Chebyshev expansion are required to accurately simulate electron dynamics. In particular, for the H$_2$ molecule, we truncated the series at 120 terms to be contrasted to a much larger value of 2000 for the spectral function calculation. 
We analyzed the performance of the algorithm in the presence of sampling noise, assuming that the expectation values of the individual terms in Chebyshev series can be computed without systematic error. This implies that at $k$-th step of recursion, the Chebyshev vector $\ket{\chi_k}$ can be prepared by a compact and expressive ansatz. We anticipate the number of parameters in such an ansatz to increase with $k$ making cost function optimization increasingly challenging due to the notorious barren plateau phenomenon~\cite{mcclean_barren_2018}. To some extent, it can be mitigated by adaptive ansatz construction~\cite{grimsley_adapt-vqe_2022} which appears to be well suited for RVSE. 
In experiments on noisy quantum devices, systematic errors are unavoidable due to the presence of decoherence effects. Similar to the sampling noise, hardware errors can get amplified throughout recursion deteriorating the accuracy of the higher order terms. We leave a more detailed exploration of this effect for future work. 
Another direction is to explore the performance of RSVE applied to expectation values of other functions of the Hamiltonian. One interesting example is the threshold function from Dong, Tong, and Lin~\cite{dong2022ground}, which is used for ground state energy estimation. Imaginary time evolution is another promising application for RVSE~\cite{kosloff_direct_1986}. 
Finally, RVSE is fully compatible with the KPM framework in that it is not limited to the Dirichlet kernel adopted in this work. Furthermore, it can be incorporated into hybrid KPM methods~\cite{irfan_hybrid_2019} and used in conjunction with embedding techniques~\cite{booth_spectral_2015} to study spectral properties of condensed matter systems. As such, we hope it opens new avenues of research directed at identifying quantum computing applications holding a promise for industrially relevant quantum advantage.



\section{Acknowledgement}

We thank Chong Sun, Max Radin, and Jerome Gonthier for detailed feedback on the manuscript. We thank Lasse Bj\o rn Kristensen for helpful discussions. P.W.K.J acknowledges support by Augustinus Fonden, Knud H\o{}jgaards Fonden, and Viet-Jacobsens Fonden. For the simulations, OpenFermion~\cite{mcclean_openfermion_2020} and Psi4~\cite{turney_psi4_2011} were used.

\appendix

\section{Error Propagating in the Chebyshev Expansion}
\label{app_cheby}

\subsection{Error scaling}
\label{subsec:var}

Consider an operator $\hat{\mathcal{O}}$ represented as a sum of Pauli strings $\hat{\mathcal{O}} = \sum_{\mathbf{k}} c_{\mathbf{k}} \hat{P}_{\mathbf{k}}$, where $\hat{P}_{\mathbf{k}} = \bigotimes_{i=1}^N \hat{\sigma}_{k_i}^{(i)}$, $\hat{\sigma}_{k_i}^{(i)}\in \{ I, \hat{\sigma}_x^{(i)}, \hat{\sigma}_y^{(i)}, \hat{\sigma}_z^{(i)} \}$. The expectation value of $\hat{\mathcal{O}}$ with respect to an \emph{N}-qubit quantum state $|\psi\rangle$ is a sum of the expectation values of the Pauli strings, $\langle \hat{\mathcal{O}} \rangle = \langle \psi | \hat{\mathcal{O}}| \psi \rangle = \sum_{\mathbf{k}} c_{\mathbf{k}} \langle \hat{P}_{\mathbf{k}} \rangle$. Let us introduce individual estimators for each $\langle \hat{P}_{\mathbf{k}} \rangle$, $\Pi_{\mathbf{k}}$. $\Pi_{\mathbf{k}}$ is a random variable defined as an average of $S_{\mathbf{k}}$ independent and identically distributed Bernoulli random variables $\xi_l$, each encoding a Pauli measurement outcome $\{-1, +1\}$, i.e. $\Pi_{\mathbf{k}} = \frac{1}{S_{\mathbf{k}}} \sum_{l = 1}^{S_{\mathbf{k}}} \xi_l $. The probability for $\xi_l$ to be $\pm 1$ is calculated based on the postulates of quantum mechanics

\begin{align}
    \text{Pr}(\xi_l = s) = \langle \psi | \hat{ \mathcal{P}}_s | \psi \rangle,
\end{align}
where $\hat{ \mathcal{P}}_s$ is the orthogonal projector on the eigenspace of $\hat{P}_{\mathbf{k}}$ with the eigenvalue $s$. Since the set of eigenvectors of $\hat{P}_{\mathbf{k}}$ is complete, $\hat{ \mathcal{P}}^{({\mathbf{k}})}_1 + \hat{ \mathcal{P}}^{({\mathbf{k}})}_{-1} = 1$. Also, $\hat{P}_{\mathbf{k}} = \hat{ \mathcal{P}}^{({\mathbf{k}})}_1 - \hat{ \mathcal{P}}^{({\mathbf{k}})}_{-1}$ (spectral decomposition). Therefore, $ \hat{ \mathcal{P}}^{({\mathbf{k}})}_{\pm 1} = \frac{1 \pm \hat{P}_{\mathbf{k}}}{2}$. It follows

\begin{align}
    \text{Pr}(\xi_l = \pm 1) =\frac{1 \pm \langle \hat{P}_{\mathbf{k}}\rangle}{2}.
\end{align}

An explicit expression for the probability allows computing the expectation value and the variance of $\xi_l$, 

\begin{align}
\langle \xi_l \rangle &=  Pr(\xi_l = +1) -  Pr(\xi_l = -1) \\
&= \frac{1 + \langle \hat{P}_{\mathbf{k}}\rangle}{2} -  \frac{1 - \langle \hat{P}_{\mathbf{k}}\rangle}{2} \\
&= \langle \hat{P}_{\mathbf{k}}\rangle \\
\operatorname{Var}(\xi_l) &= 1 - \langle \hat{P}_{\mathbf{k}}\rangle^2 \\
\end{align}
Similarly,  $\langle \Pi_{\mathbf{k}} \rangle = \langle \hat{P}_{\mathbf{k}}\rangle$ and 

\begin{align}
\operatorname{Var}(\Pi_{\mathbf{k}}) &= \operatorname{Var}\bigg(\frac{1}{S_{\mathbf{k}}} \sum_{l = 1}^{S_{\mathbf{k}}} \xi_l \bigg)\\
&= \frac{1}{S^2_{\mathbf{k}}} \sum_{l = 1}^{S_{\mathbf{k}}}\operatorname{Var}( \xi_l)\\
&= (1 - \langle \hat{P}_{\mathbf{k}}\rangle^2) / S_{\mathbf{k}}    
\end{align}
From the equation for $\langle \Pi_{\mathbf{k}} \rangle$ , we see that $E = \sum_{\mathbf{k}} c_{\mathbf{k}} \Pi_{\mathbf{k}}$ is an unbiased estimator for $\langle \hat{\mathcal{O}} \rangle$. Indeed,

\begin{align}
    \langle E \rangle &= \sum_{\mathbf{k}} c_{\mathbf{k}} \langle \Pi_{\mathbf{k}} \rangle\\
    &= \sum_{\mathbf{k}} c_{\mathbf{k}} \langle \hat{P}_{\mathbf{k}} \rangle\\
    &= \langle \hat{\mathcal{O}} \rangle.
\end{align}
Since Cov$(\Pi_{\mathbf{k}}, \Pi_{\mathbf{l}}) = \delta_{\mathbf{k}\mathbf{l}} \operatorname{Var}(\Pi_{\mathbf{k}})$,

\begin{align}
    \operatorname{Var}\left(E\right) &= \sum_{\mathbf{k}}\left|c_{\mathbf{k}}\right|^2 \operatorname{Var}\left(\Pi_{\mathbf{k}}\right)\\
    &= \sum_{\mathbf{k}}\left|c_{\mathbf{k}}\right|^2 \operatorname{Var}\left(\hat{P}_{\mathbf{k}}\right) / S_{\mathbf{k}} \\
  &\leq \sum_{\mathbf{k}} \frac{\left|c_{\mathbf{k}}\right|^2}{S_{\mathbf{k}}}. \label{eq_app_var_final}
\end{align}

\subsection{Error Propagating in the Normalizing Constants}
\label{subsec:norm_constant}

We compute the expectation value of the operator $\hat{F}(\hat{H}|E)$ by expanding the operator in a \emph{K}-term Chebyshev polynomials expansion as 

\begin{align}
\braket{\chi_0|\hat{F}_K(\hat{H}|E)|\chi_0} 
&=  \braket{\chi_0|\sum^{K}_{k=0} c_k(E)\hat{T}_k(\hat{H})|\chi_0} \\
&= \sum^{K}_{k=0} c_k(E) \bra{\chi_0}\hat{T}_k(\hat{H}) \ket{\chi_0}  \\
&= \sum^K_{k = 0} c_k(E) \norm{\ket{\chi_0}} \norm{\ket{\chi_k}} \braket{\boldsymbol{0}|\hat{U}^\dagger(\boldsymbol{\theta}_0)\hat{U}(\boldsymbol{\theta}_k)|\boldsymbol{0}},\label{eq_F_exp}
\end{align}
where $\ket{\boldsymbol{0}} \equiv \ket{0\hdots 0}$, $\hat{H}$ is the system Hamiltonian, and $c_k(E) $ is the expansion coefficients which can be obtained from Eq. \eqref{eq:expand_coefs}. The wave function overlap $\braket{\boldsymbol{0}|\hat{U}^\dagger(\boldsymbol{\theta}_0)\hat{U}(\boldsymbol{\theta}_k)|\boldsymbol{0}}$ can be evaluated on a quantum computer using the Hadamard test circuit.  We compute the normalizing constants $\norm{\ket{\chi_k}}$  and angles $ \boldsymbol{\theta}_k$ by optimizing the cost function $F_k(\boldsymbol{\theta})$, Eq.~\eqref{eq:overlap:func}, and for $\boldsymbol{\theta} = \boldsymbol{\theta}_k$ we obtain 

\begin{align}
F_k(\boldsymbol{\theta}_k) = \norm{\ket{\chi_k}}. \label{eq_norm_1}
\end{align}
In this section, we will investigate how the error propagates through the normalizing constants when sampling error is introduced in \eqref{eq_norm_1}. We will assume the ansatz is expressive enough to reach the target states at each iteration, and the optimization algorithm is able to locate the global maximum at each iteration, i.e., the optimal angles $\boldsymbol{\theta}_k$ are obtained at each iteration.

We define:

\begin{align}
\mu_{\textbf{k}}^{(r,s)} &= \braket{\boldsymbol{0}| \hat{U}^{\dagger}(\boldsymbol{\theta}_r) \hat{P}_{\textbf{k}} \hat{U}(\boldsymbol{\theta}_{s})|\boldsymbol{0} } \label{eq_app_mu}
\\[0.3cm]
\nu^{(r,s)} &= \braket{\boldsymbol{0}| \hat{U}^{\dagger}(\boldsymbol{\theta}_r)  \hat{U}(\boldsymbol{\theta}_{s})|\boldsymbol{0}}, \label{eq_app_nu}
\end{align}
and the cost function can then be written as

\begin{align}
F_k(\boldsymbol{\theta}_k)= \norm{\ket{\chi_k}} = \bigg|2~ \norm{\ket{\chi_{k-1}}} \sum_{\textbf{k}} c_{\textbf{k}}\mu_{\textbf{k}}^{(k,k-1)} -   \norm{\ket{\chi_{k-2}}} \nu^{(k,k-2)}\bigg|.  \label{eq_fk_new}
\end{align}

If the initial state is normalized then $\norm{\ket{\chi_0}} = 1$. Otherwise, we would need to compute the initial normalizing constant.

For $ k = 1$, the cost function reads

\begin{align}
 F_1(\boldsymbol{\theta}_1) &= \norm{\ket{\chi_1}} \\ &= \bigg|\norm{\ket{\chi_0}}\sum_{\textbf{k}} c_{\textbf{k}} \mu_{\textbf{k}}^{(1,0)} \bigg| \\
 &=  \bigg|\norm{\ket{\chi_0}}\sum_{\textbf{k}} c_{\textbf{k}} \text{Re}[\mu_{\textbf{k}}^{(1,0)}] +i\norm{\ket{\chi_0}}\sum_{\textbf{k}} c_{\textbf{k}} \text{Im}[\mu_{\textbf{k}}^{(1,0)}]\bigg|.\label{eq_F1}
\end{align}
Note that the Hamiltonian coefficients $c_\textbf{k}$ are real. Let $\hat{\mu}^{(r,s)}_\textbf{k}$ be our estimate of $\mu^{(r,s)}_\textbf{k}$ from the Hadamard test evaluations, i.e., 

\begin{align}
\hat{\mu}_\textbf{k}^{(r,s)} &=\frac{1}{\nmea} \bigg(\sum^\nmea_{l =  1} {}^{\text{Re}}(\xi^{(r,s)}_\textbf{k})_{(l)} + i  \sum^\nmea_{l = 1} {}^{\text{Im}}(\xi^{(r,s)}_\textbf{k})_{(l)}  \bigg)
\end{align}
where $(\xi^{(r,s)}_\textbf{k})_{(l)} \in \{-1,1\}$ is the \emph{l}-th measurement outcome, Re (Im) indicates the output of the Hadamard test circuit when computing the real (imaginary) part of the wave function overlap which correspond to slightly different circuits, and $\nmea$ is the number of measurements used to evaluate the wave function overlap. The variance of the wave function overlap estimators  used within Hadamard test can be bounded  as (see Appendix \ref{subsec:var})

\begin{align}
&\operatorname{Var}\bigg(  \norm{\ket{\chi_0}}\sum_{\textbf{k}} c_{\textbf{k}} \text{Re}[\hat{\mu}_{\textbf{k}}^{(1,0)}] \bigg) \leq \frac{1}{\nmea}\norm{\ket{\chi_0}}^2\sum_{\textbf{k}} \abs{c_{\textbf{k}}}^2  \label{eq_app_var_bound} \\
&\operatorname{Var}\bigg(i\norm{\ket{\chi_0}}\sum_{\textbf{k}} c_{\textbf{k}} \text{Im}[\hat{\mu}_{\textbf{k}}^{(1,0)}]\bigg)  \leq \frac{1}{\nmea}\norm{\ket{\chi_0}}^2\sum_{\textbf{k}} \abs{c_{\textbf{k}}}^2.  \label{eq_app_var_bound_1}
\end{align}
Considering the large sample limit such that the estimator $\hat{X}$ can be written as a random number sampled from $\hat{X} \leftarrow \sqrt{\operatorname{Var}(\hat{X})}\mathcal{N}(0,1) + \mathbb{E}[\hat{X}]$, where $\mathcal{N}(0,1)$ is a random distribution with zero mean and variance equal to one. In that limit, we can write the  estimator  of the cost function $\hat{F}_1(\boldsymbol{\theta}_1)$ as a (biased) random variable

\begin{align}
\hat{F}_1(\boldsymbol{\theta}_1) =  \left| \norm{\ket{\chi_0}}\sum_{\textbf{k}} c_{\textbf{k}} \mu_{\textbf{k}}^{(1,0)}+  \phi_{\mu,1} + i\phi_{\mu,1} \right| \leq F_1(\boldsymbol{\theta}_1) + \left| \phi_{\mu,1} + i\phi_{\mu,1} \right| \label{eq_app_upper_cost}
\end{align}
since $\hat{\mu}^{(1,0)}_{\textbf{k}}$ is an unbiased estimator, $\mathbb{E}[\hat{\mu}^{(1,0)}_{\textbf{k}}] = \mu^{(1,0)}_{\textbf{k}}$, and $\phi_{\mu,1} $ is a random number with zero mean sampled from the distribution

\begin{align}
\phi_{\mu,1} \leftarrow \sqrt{\frac{1}{\nmea}\norm{\ket{\chi_0}}^2\sum_{\textbf{k}} \abs{c_{\textbf{k}}}^2 } ~ \mathcal{N}(0,1).
\end{align}
Here $ x \leftarrow  \mathcal{N}$ indicates \emph{x} is randomly selected from the normal distribution $\mathcal{N}$. The variance and the expected value of the estimator, Eq. \eqref{eq_app_upper_cost}, can be approximated as 

\begin{align}
\operatorname{Var}\left(\hat{F}_1(\boldsymbol{\theta}_1) \right) &= \operatorname{Var}\bigg( F_1(\boldsymbol{\theta}_1) + \left| \phi_{\mu,1} + i\phi_{\mu,1} \right| \bigg) = \operatorname{Var}\left(\left| \phi_{\mu,1} + i\phi_{\mu,1} \right| \right) \\
\mathbb{E}\left(\hat{F}_1(\boldsymbol{\theta}_1) \right) &=\mathbb{E}\bigg( F_1(\boldsymbol{\theta}_1) + \left| \phi_{\mu,1} + i\phi_{\mu,1} \right| \bigg) =   \norm{\ket{\chi_1}}  + \mathbb{E}\left(\left| \phi_{\mu,1} + i\phi_{\mu,1} \right| \right).
 \label{eq_app_chi_1}\end{align}
For $ k = 2$, the cost function reads

\begin{align}
 F_2(\boldsymbol{\theta}_2)&=  \norm{\ket{\chi_2}} \\
 &=\left|2 \norm{\ket{\chi_{1}}} \sum_{\textbf{k}} c_{\textbf{k}} \mu_{\textbf{k}}^{(2,1)} -  \norm{\ket{\chi_{0}}} \nu^{(2,0)} \right| \\
&= \left| 2 \norm{\ket{\chi_{1}}} \sum_{\textbf{k}} c_{\textbf{k}} \bigg( \text{Re}[\mu_{\textbf{k}}^{(2,1)}] + i \text{Im}[\mu_{\textbf{k}}^{(2,1)}] \bigg) -  \norm{\ket{\chi_{0}}} \bigg( \text{Re}[\nu^{(2,0)}] + i \text{Im}[\nu^{(2,0)}] \bigg) \right|. \label{eq_F2}
\end{align}
We approximate $\norm{\ket{\chi_1}}$ with Eq.~\eqref{eq_app_chi_1}. Considering the large sample limit, and  $ \mathbb{E}\left(\left| \phi_{\mu,1} + i\phi_{\mu,1} \right| \right) \in \mathbb{R}_+$, the variance used within Hadamard test is

\begin{align}
\operatorname{Var}\left( 2\mathbb{E}\left(\hat{F}_1(\boldsymbol{\theta}_1) \right) \sum_{\textbf{k}} c_{\textbf{k}} \text{Re}[\hat{\mu}_{\textbf{k}}^{(2,1)}] \right) &\leq \frac{4 \left(\norm{\ket{\chi_1}} + \mathbb{E}\left(\left| \phi_{\mu,1} + i\phi_{\mu,1} \right| \right) \right)^2}{S} \sum_{\textbf{k}} |c_{\textbf{k}}|^2 \\
\operatorname{Var}\left( i2\mathbb{E}\left(\hat{F}_1(\boldsymbol{\theta}_1) \right) \sum_{\textbf{k}} c_{\textbf{k}} \text{Im}[\hat{\mu}_{\textbf{k}}^{(2,1)}] \right) &\leq \frac{4 \left(\norm{\ket{\chi_1}} + \mathbb{E}\left(\left| \phi_{\mu,1} + i\phi_{\mu,1} \right| \right) \right)^2}{S} \sum_{\textbf{k}} |c_{\textbf{k}}|^2 \\
\operatorname{Var}\left(- \norm{\ket{\chi_0}} \text{Re}[\hat{\nu}_{\textbf{k}}^{(2,0)}] \right)  &\leq \frac{ \norm{\ket{\chi_0}}^2}{S} \\
\operatorname{Var}\left(-i \norm{\ket{\chi_0}} \text{Im}[\hat{\nu}_{\textbf{k}}^{(2,0)}] \right) &\leq \frac{ \norm{\ket{\chi_0}}^2}{S}.
\end{align}
Considering the large sample limit, we can write the  estimator  of the cost function $\hat{F}_2(\boldsymbol{\theta}_2)$ as a biased random variable

\begin{align}
\hat{ F}_2(\boldsymbol{\theta}_2) &= \left|2 \norm{\ket{\chi_{1}}} \sum_{\textbf{k}} c_{\textbf{k}} \mu_{\textbf{k}}^{(2,1)} -  \norm{\ket{\chi_{0}}} \nu^{(2,0)}  + \phi_{\mu,2} + i \phi_{\mu,2}  + \phi_{\nu,2} + i\phi_{\nu,2} \right| \\
&\leq   F_2(\boldsymbol{\theta}_2) +  \left|  \phi_{\mu,2} + i \phi_{\mu,2}  + \phi_{\nu,2} + i\phi_{\nu,2}  \right| \label{eq_f2_esti}
\end{align}
where $\phi_{\mu,2}$ and $\phi_{\nu,2}$ are random numbers sampled from the distributions 

\begin{align}
\phi_{\mu,2}  &\leftarrow \sqrt{\frac{4 \left(\norm{\ket{\chi_1}} +  \mathbb{E}\left(\left| \phi_{\mu,1} + i\phi_{\mu,1} \right| \right) \right)^2}{S} \sum_{\textbf{k}} |c_{\textbf{k}}|^2 } ~ \mathcal{N}(0,1) \\
\phi_{\nu,2}  &\leftarrow \sqrt{\frac{\norm{\ket{\chi_0}}^2}{\nmea}} ~ \mathcal{N}(0,1).
\end{align}
The variance and the expected value of the estimator, Eq. \eqref{eq_f2_esti}, can be approximated as

\begin{align}
\operatorname{Var}\left(\hat{F}_2(\boldsymbol{\theta})_2 \right) &= \operatorname{Var}\bigg( F_2(\boldsymbol{\theta}_2) +  \left|  \phi_{\mu,2} + i \phi_{\mu,2}  + \phi_{\nu,2} + i\phi_{\nu,2}  \right| \bigg) \nonumber \\
&= \operatorname{Var}\left( \left|  \phi_{\mu,2} + i \phi_{\mu,2}  + \phi_{\nu,2} + i\phi_{\nu,2}  \right|\right) \\
\mathbb{E}\left(\hat{F}_2(\boldsymbol{\theta}_2) \right) &=\mathbb{E}\bigg( F_2(\boldsymbol{\theta}_2) +  \left|  \phi_{\mu,2} + i \phi_{\mu,2}  + \phi_{\nu,2} + i\phi_{\nu,2}  \right| \bigg) \\
&=   \norm{\ket{\chi_2}} + \mathbb{E}\left( \left|  \phi_{\mu,2} + i \phi_{\mu,2}  + \phi_{\nu,2} + i\phi_{\nu,2}  \right| \right)
\end{align}
In general, we can write 

\begin{align}
\norm{\ket{\chi_k}}_{\text{noisy}} = \norm{\ket{\chi_k}} + \epsilon_k \label{eq_app:noisy_k}
\end{align}
where $\norm{\ket{\chi_k}}_{\text{noisy}} = \mathbb{E}\left(\hat{F}_k(\boldsymbol{\theta}_k) \right) $,  $ \epsilon_k = \mathbb{E}\left( \left|  \phi_{\mu,k} + i \phi_{\mu,k}  + \phi_{\nu,k} + i\phi_{\nu,k}  \right|\right) $, and the $\phi$'s are random numbers selected from the distributions

\begin{align}
\phi_{\mu,0} &= 0 \\
\phi_{\mu,1} &\leftarrow \sqrt{\frac{\norm{\ket{\chi_0}}^2 }{\nmea} \sum_{\textbf{k}} \abs{c_{\textbf{k}}}^2} ~ \mathcal{N}(0,1) \\
\phi_{\mu,k}&\leftarrow \sqrt{\frac{4 \left(\norm{\ket{\chi_{k-1}}} + \mathbb{E}\left( \left|  \phi_{\mu,k-1} + i \phi_{\mu,k-1}  + \phi_{\nu,k-1} + i\phi_{\nu,k-1}  \right| \right)\right)^2}{S} \sum_{\textbf{k}} |c_{\textbf{k}}|^2 } ~ \mathcal{N}(0,1)
\end{align}
and

\begin{align}
\phi_{\nu,0} &= 0 \\
\phi_{\nu,1} &= 0 \\
\phi_{\nu,k}&\leftarrow \sqrt{\frac{\left(\norm{\ket{\chi_{k-2}}} + \mathbb{E}\left( \left|  \phi_{\mu,k-2} + i \phi_{\mu,k-2}  + \phi_{\nu,k-2} + i\phi_{\nu,k-2}  \right| \right)\right)^2}{\nmea}} ~ \mathcal{N}(0,1).
\end{align}

\subsection{Error Propagating in the Expectation Value}
\label{subsec:exp}
Considering the large sample limit, the noisy expectation value takes the form 

\begin{align}
\braket{\chi_0|\hat{F}_K(\hat{H}|E)|\chi_0}  &\approx\sum^K_{k = 0} c_k(E) \norm{\ket{\chi_0}} \norm{\ket{\chi_k}}_{\text{noisy}}  \hat{\nu}^{(0,k)}  \nonumber \\
&= \sum^K_{k = 0} c_k(E) \norm{\ket{\chi_0}} \left(\norm{\ket{\chi_k}} + \epsilon_k \right) \left(  \nu^{(0,k)} + \phi_k\right) \label{eq_app:noisy_exp}
\end{align}
where $\hat{\nu}^{(0,k)} $ is our estimator of the true overlap in Eq. \eqref{eq_app_nu} which, in the large sample limit, can be written as a unbiased random variable with $\phi_k$ being a random number sampled from the distribution 

\begin{align}
\phi_k \leftarrow  \frac{1}{S}   \mathcal{N}(0,1)
\end{align}
where \emph{S} is the number of measurements and $ \mathcal{N}(0,1)$ is a random distribution with zero mean and variance equal to one. The expectation value of \eqref{eq_app:noisy_exp} reads

\begin{align}
&\mathbb{E}\left[ \sum^K_{k = 0} c_k(E) \norm{\ket{\chi_0}} \left(\norm{\ket{\chi_k}} + \epsilon_k \right) \left(  \nu^{(0,k)} + \phi_k\right)  \right] \nonumber \\
&= \braket{\chi_0|\hat{F}_K(\hat{H}|E)|\chi_0} + \mathbb{E}\left[ \sum^K_{k = 0} c_k(E) \norm{\ket{\chi_0}}  \left( \norm{\ket{\chi_k}} \phi_k+ \epsilon_k\nu^{(0,k)}  + \epsilon_k \phi_k\right)  \right]  \\
&= \braket{\chi_0|\hat{F}_K(\hat{H}|E)|\chi_0} +  \sum^K_{k = 0} c_k(E) \norm{\ket{\chi_0}} \epsilon_k\nu^{(0,k)} 
\end{align}
since $\mathbb{E}\left[\phi_k \right] = 0$. Thus, in the large sample limit, the absolute error in the expectation value can be written as 

\begin{align}
\Delta = \left|\sum^K_{k = 0} c_k(E) \norm{\ket{\chi_0}} \epsilon_k\nu^{(0,k)}\right|.
\end{align}
Note that the error in the expectation value is almost identical to the Chebyshev expansion, but $\norm{\ket{\chi_k}} \rightarrow \epsilon_k$. 

\bibliographystyle{apsrev4-2}
\bibliography{refs}

\end{document}